\documentclass[reprint, aps,nofootinbib]{revtex4-2}
\usepackage{amssymb}
\usepackage{amsfonts}
\usepackage{amsmath}
\usepackage{graphicx}
\usepackage{dcolumn}
\usepackage{bm}
\usepackage{tikz}
\usepackage{tikz-3dplot}
\usetikzlibrary{decorations.pathmorphing}
\tikzset{
    partial ellipse/.style args={#1:#2:#3}{
        insert path={+ (#1:#3) arc (#1:#2:#3)}
    }
}
\tikzset{
    mline/.pic={
        \draw (0,-3) -- (0,3);
        \draw (1.04,0) [partial ellipse=0:360:0.7 and 1.5];
        \draw (-1.04,0) [partial ellipse=0:360:0.7 and 1.5];
        \draw (1.5,0) [partial ellipse={0}:{360}:1.3 and 2];
        \draw (-1.5,0) [partial ellipse={0}:{360}:1.3 and 2];    
        \draw (3,0) [partial ellipse={90+80.5-20}:{360-90-80.5+20}:2.9 and 6];
        \draw (-3,0) [partial ellipse={90-80.5+20}:{-90+80.5-20}:2.9 and 6];
        \draw (2.14,0) [partial ellipse={90+71-50}:{360-90-71+50}:2 and 3];
        \draw (-2.14,0) [partial ellipse={90-71+50}:{-90+71-50}:2 and 3];
    }
}
\tikzset{
    pulsar/.pic={
        \pic[rotate=-60,transform shape] {mline};
        \shadedraw[ball color = gray!0] (0,0) circle (1);
        \draw[thick] (0,1) -- (0,3.5);
        \draw[thick] (0,-1) -- (0,-3.5);
        \draw[stealth-] (0,3.5) [partial ellipse={70}:{-250}:0.5 and 0.2];
    }
}
\tikzset{
    dish/.pic={
        \shadedraw[right color = gray, left color = white, shading = axis] (0,0) ellipse (1 and 2);
            \draw[thick] (0,-1.5) -- (1.5,0);
            \draw[thick] (0.5,0.1) -- (1.5,0);
            \draw[thick] (0,1.5) -- (1.5,0);
            \draw[thick] (-0.5,-0.1) -- (1.5,0);
    }
}
\tikzset{
    telescope/.pic={
        \draw (-0.3,0) -- (-0.6,-2.5);
        \draw (0.3,0) -- (0.6,-2.5);
        \draw (0.6,-2.5) arc (0:-180:0.6 and 0.2);
        \pic[rotate=30] {dish};
    }
}
\tikzset{snake it/.style={decorate, decoration=snake}}
\tikzset{
    emgw/.pic={
        \pic[scale=0.6] at (4,{4/sqrt(3)}) {pulsar};
        \pic at (-4,{-4/sqrt(3)}) {telescope};
        \draw[thick,-latex] (2,{2/sqrt(3)}) -- (-2,{-2/sqrt(3)}) node[very near end,above,yshift=5] {EMW};
        \draw[snake it] (-2,{2/sqrt(3)}) -- (2,{-2/sqrt(3)}) node[very near end,above,yshift=5] {GW};
    }
}

\begin{document}

\title{Detection of gravitational waves by light perturbation}
\author{Dong-Hoon Kim}
\email{ki13130@gmail.com}
\affiliation{Department of Physics and Astronomy, Seoul National University,
Seoul 08826, Republic of Korea}
\author{Chan Park}
\email{iamparkchan@gmail.com}
\affiliation{National Institute for Mathematical Sciences, Daejeon 34047,
Republic of Korea}
\date{\today }

\begin{abstract}
Light undergoes perturbation as gravitational waves pass by. This is shown
by solving Maxwell's equations in a spacetime with gravitational waves; a
solution exhibits a perturbation due to gravitational waves. We determine
the perturbation for a general case of both light and gravitational waves
propagating in arbitrary directions. It is also shown that a perturbation of
light due to gravitational waves leads to a delay of the photon transit
time, which implies an equivalence between the perturbation analysis of
Maxwell's equations and the null geodesic analysis for photon propagation.
We present an example of application of this principle with regard to the
detection of gravitational waves via a pulsar timing array, wherein our
perturbation analysis for the general case is employed to show how the
detector response varies with the incident angle of a light pulse with
respect to the detector.
\end{abstract}

\maketitle

\preprint{APS/123-QED}

\section{Introduction\label{intro}}

Light is the most common and important tool in astronomy, due to its
property of carrying energy and information about its sources: it can reach
an observer even at quite a distance, providing clues about astronomical
sources responsible for its creation. Artificially created light is also in
use in astronomy; e.g., laser light being commonly used for interferometry.
Laser interferometers exploit another prominent and interesting property of
light to detect gravitational waves (GWs): its interaction with other waves
-- GWs. However, from a perspective based on general relativity, this
interaction can be viewed as a perturbation of light due to GWs; that is,
light is perturbed as GWs pass through space in which it travels.

There is considerable significance in studying the aforesaid property of
light in regard to the detection of GWs; e.g., by means of laser
interferometers -- LIGO, VIRGO, GEO600, KAGRA, LIGO-India, eLISA, etc. \cite%
{Abbott,Grote,Somiya,Iyer,AS} or pulsar timing arrays (PTAs) -- EPTA, PPTA,
IPTA, SKA, etc. \cite{Kramer,Hobbs,Manchester,Dewdney}. In principle, in all
these detection schemes, we utilize the delay of the photon transit time due
to GWs; that is, the effect resulting from a photon that undergoes deviation
from its straight path while propagating in a spacetime with GWs. But this
effect can be shown to be equivalent to the perturbation of an
electromagnetic field due to GWs, as a direct consequence of a solution of
Maxwell's equations in the spacetime perturbed by GWs.

There were numerous studies about electrodynamics in a GW background based
on Maxwell's equations. Among others, Calura and Montanari \cite{Calura}
solved Maxwell's equations only in the framework of the linearized general
relativity and provided the exact solution to the problem, expressed like
the Fourier-integral, by considering a general case for the GW frequency,
rather than using the \textit{geometrical-optics} approximation. Hacyan \cite%
{Hacyan1,Hacyan2} analyzed the interaction of electromagnetic waves with a
plane-fronted GW and derived the corresponding formulas for Stokes
parameters and the rotation angle of polarization. Cabral and Lobo \cite%
{Cabral} obtained electromagnetic field oscillations induced by GWs and
found that these lead to the presence of longitudinal modes and dynamical
polarization patterns of electromagnetic radiation.

In this paper, we address the issue how light is perturbed in the presence
of GWs from a general relativistic perspective. In the context of the
geometrical-optics approach, we solve Maxwell's equations for a general
configuration, wherein both light and GWs are assumed to propagate in
arbitrary directions. Unlike the aforementioned related works in the
literature, in which rather simple configurations with regard to the propagation directions of light and GWs are
assumed, our investigation aims to achieve full generality in this regard,
thereby providing practical results that can be readily used for analyzing
various detection schemes for GWs, wherein such generality might be
required; e.g., in a PTA, light pulses from different pulsars may arrive at
one detector with various incident angles with respect to the stationary
reference frame (or detector frame), while GWs may come from arbitrary
directions to cross them. Largely, the paper proceeds in three steps through
Sects. \ref{maxwell}-\ref{app} as follows. In Sect. \ref{maxwell}, we solve
Maxwell's equations in a spacetime perturbed by GWs for the general
configuration between light and GWs, presenting a solution to first order in
the strain amplitude $h$. In Sect. \ref{pert}, we establish the equivalence
between a perturbation of light due to GWs and a delay of the photon transit
time, with the former implied from the solution of the Maxwell's equations
obtained in Sect. \ref{maxwell} and the latter implied from the null
geodesic of photon propagation. In Sect. \ref{app}, application of our
analysis from Sect. \ref{pert} to the detection of GWs via a PTA is
discussed. Using the equivalence in the context of the general configuration
between light and GWs, we determine the detector response function for light
pulses incident on a detector at various angles.

\section{Analysis and Results\label{form}}

\subsection{Solving Maxwell's equations for light perturbed by GWs\label%
{maxwell}}

What happens to light when GWs pass through space in which it propagates?
This can be answered by solving Maxwell's equations defined in a spacetime
perturbed by GWs. For example, an electromagnetic field as a solution to the
Maxwell's equations can describe a light ray from a star or a laser beam in
an interferometer being perturbed by GWs. For simplicity, we consider a case
of a \textit{monochromatic} electromagnetic wave (EMW) perturbed by \textit{%
monochromatic} GWs. However, for the sake of generality of the
configuration, we assume that both light and GWs propagate in arbitrary
directions. Our analysis follows.

Suppose that GWs propagate along the $z^{\prime }$-axis while being
polarized in the $x^{\prime }y^{\prime }$-plane in a quadrupole manner:%
\begin{eqnarray}
h_{ij}^{+} &=&h_{+}\left( e_{i}^{x^{\prime }}\otimes e_{j}^{x^{\prime
}}-e_{i}^{y^{\prime }}\otimes e_{j}^{y^{\prime }}\right) e^{\mathrm{i}\left(
kz^{\prime }-\omega _{\mathrm{g}}t\right) },  \label{hp} \\
h_{ij}^{\times } &=&h_{\times }\left( e_{i}^{x^{\prime }}\otimes
e_{j}^{y^{\prime }}+e_{i}^{y^{\prime }}\otimes e_{j}^{x^{\prime }}\right) e^{%
\mathrm{i}\left( kz^{\prime }-\omega _{\mathrm{g}}t-\pi /2\right) },
\label{hc}
\end{eqnarray}%
where $i$, $j$ refer to the coordinates $\left( x^{\prime },y^{\prime
},z^{\prime }\right) $, and $h_{+}$ and $h_{\times }$ represent the strain
amplitude for $+$ and $\times $ polarization states, respectively, and $%
\omega _{\mathrm{g}}$ denotes the GW frequency; $\omega _{\mathrm{g}}=ck$
with $c$ being the speed of light and $k$ being the wavenumber for GW. Then
the spacetime geometry reads in the coordinates $\left( t,x^{\prime
},y^{\prime },z^{\prime }\right) $:%
\begin{eqnarray}
ds^{2} &=&-c^{2}dt^{2}+\left[ 1+\Re \left( h_{+}e^{\mathrm{i}\left(
kz^{\prime }-\omega _{\mathrm{g}}t\right) }\right) \right] dx^{\prime 2} 
\notag \\
&&+2\Re \left( h_{\times }e^{\mathrm{i}\left( kz^{\prime }-\omega _{\mathrm{g%
}}t-\pi /2\right) }\right) dx^{\prime }dy^{\prime }  \notag \\
&&+\left[ 1-\Re \left( h_{+}e^{\mathrm{i}\left( kz^{\prime }-\omega _{%
\mathrm{g}}t\right) }\right) \right] dy^{\prime 2}+dz^{\prime 2}.  \label{ds}
\end{eqnarray}

However, one can consider the coordinates $\mathbf{x}^{\prime }\equiv \left(
x^{\prime },y^{\prime },z^{\prime }\right) $ as rotated from the coordinates 
$\mathbf{x}\equiv \left( x,y,z\right) $ through Euler angles $\left\{ \phi
,\theta ,\psi \right\} $ \cite{Goldstein,Rakhmanov}:%
\begin{equation}
\mathbf{x}^{\prime }=\mathbf{R}\left( \phi ,\theta ,\psi \right) \mathbf{x,}
\label{eu}
\end{equation}%
where we let $\mathbf{x}^{\prime }$ and $\mathbf{x}$ refer to the
coordinates in the \textit{GW} frame and the \textit{detector} frame,
respectively (see Fig.~\ref{fig0} for illustration), and 
\begin{equation}
\mathbf{R}\left( \phi ,\theta ,\psi \right) =\mathbf{R}_{3}\left( \psi
\right) \mathbf{R}_{2}\left( \theta \right) \mathbf{R}_{1}\left( \phi
\right) ,  \label{R}
\end{equation}%
with 
\begin{align}
& \mathbf{R}_{1}\equiv \left[ 
\begin{array}{ccc}
\cos \phi & \sin \phi & 0 \\ 
-\sin \phi & \cos \phi & 0 \\ 
0 & 0 & 1%
\end{array}%
\right] ,\mathbf{R}_{2}\equiv \left[ 
\begin{array}{ccc}
\cos \theta & 0 & -\sin \theta \\ 
0 & 1 & 0 \\ 
\sin \theta & 0 & \cos \theta%
\end{array}%
\right] ,  \notag \\
& \mathbf{R}_{3}\equiv \left[ 
\begin{array}{ccc}
\cos \psi & \sin \psi & 0 \\ 
-\sin \psi & \cos \psi & 0 \\ 
0 & 0 & 1%
\end{array}%
\right] ,  \label{R123}
\end{align}%
and $\left\{ \phi ,\theta \right\} $ denote the direction angles in
spherical coordinates, defined with respect to the coordinates $\left(
x,y,z\right) $, and $\psi $ denotes the polarization-ellipse angle \cite{Pai}%
. Resulting from these rotations, the spacetime geometry in the coordinates $%
\left( t,x^{\prime },y^{\prime },z^{\prime }\right) $ given by (\ref{ds}) is
now rewritten in the coordinates $\left( t,x,y,z\right) $: 
\begin{align}
ds^{2}=-c^{2}dt^{2}& +\sum_{i,j=1,2,3}\left[ \delta _{ij}+\alpha _{ij}\left(
\phi ,\theta ,\psi \right) H_{+}\right.  \notag \\
& \left. +\beta _{ij}\left( \phi ,\theta ,\psi \right) H_{\times }\right]
dx_{i}dx_{j},  \label{ds1}
\end{align}%
where%
\begin{eqnarray}
\alpha _{11}\left( \phi ,\theta ,\psi \right) &=&\cos \left( 2\psi \right)
\left( \cos ^{2}\theta \cos ^{2}\phi -\sin ^{2}\phi \right)  \notag \\
&&-2\sin \left( 2\psi \right) \cos \theta \cos \phi \sin \phi ,  \notag \\
\alpha _{12}\left( \phi ,\theta ,\psi \right) &=&\cos \left( 2\psi \right)
\left( 1+\cos ^{2}\theta \right) \cos \phi \sin \phi  \notag \\
&&+\sin \left( 2\psi \right) \cos \theta \left( 2\cos ^{2}\phi -1\right) , 
\notag \\
\alpha _{13}\left( \phi ,\theta ,\psi \right) &=&-\cos \left( 2\psi \right)
\cos \theta \sin \theta \cos \phi  \notag \\
&&+\sin \left( 2\psi \right) \sin \theta \sin \phi ,  \notag \\
\alpha _{21}\left( \phi ,\theta ,\psi \right) &=&\alpha _{12}\left( \phi
,\theta ,\psi \right) ,  \notag \\
\alpha _{22}\left( \phi ,\theta ,\psi \right) &=&\cos \left( 2\psi \right)
\left( \cos ^{2}\theta \sin ^{2}\phi -\cos ^{2}\phi \right)  \notag \\
&&+2\sin \left( 2\psi \right) \cos \theta \cos \phi \sin \phi ,  \notag \\
\alpha _{23}\left( \phi ,\theta ,\psi \right) &=&-\cos \left( 2\psi \right)
\cos \theta \sin \theta \sin \phi  \notag \\
&&-\sin \left( 2\psi \right) \sin \theta \cos \phi ,  \notag \\
\alpha _{31}\left( \phi ,\theta ,\psi \right) &=&\alpha _{13}\left( \phi
,\theta ,\psi \right) ,  \notag \\
\alpha _{32}\left( \phi ,\theta ,\psi \right) &=&\alpha _{23}\left( \phi
,\theta ,\psi \right) ,  \notag \\
\alpha _{33}\left( \phi ,\theta ,\psi \right) &=&\cos \left( 2\psi \right)
\sin ^{2}\theta ,  \label{alpha}
\end{eqnarray}%
and 
\begin{equation}
\beta _{ij}\left( \phi ,\theta ,\psi \right) =\alpha _{ij}\left( \phi
,\theta ,\psi +\pi /4\right) ,  \label{beta}
\end{equation}%
and 
\begin{eqnarray}
H_{+} &\equiv &\Re \left( h_{+}\exp \left[ \mathrm{i}\left( \mathbf{k\cdot x}%
-\omega _{\mathrm{g}}t\right) \right] \right) ,  \label{Hp} \\
H_{\times } &\equiv &\Re \left( h_{+}\exp \left[ \mathrm{i}\left( \mathbf{%
k\cdot x}-\omega _{\mathrm{g}}t-\pi /2\right) \right] \right) ,  \label{Hc}
\end{eqnarray}%
with%
\begin{eqnarray}
\mathbf{k} &=&\left( k_{x},k_{y},k_{z}\right)  \notag \\
&\equiv &\left( k\sin \theta \cos \phi ,k\sin \theta \sin \phi ,k\cos \theta
\right) .  \label{k}
\end{eqnarray}%
Here one should note the following important property: the dependence on the
polarization angle $\psi $ in (\ref{alpha}) and (\ref{beta}) exhibits the
spin-2 tensor modes of the $+$ and $\times $ polarizations.

Our light perturbed by GWs can be described by Maxwell's equations defined
in curved (perturbed) spacetime as given by Eq. (\ref{ds1}): in the Lorenz
gauge \cite{MTW}, 
\begin{equation}
\square A^{\mu }-R^{\mu }{}_{\nu }A^{\nu }=0,  \label{me}
\end{equation}%
where $\square A^{\mu }\equiv g^{\nu \rho }\nabla _{\nu }\nabla _{\rho
}A^{\mu }$ means the d'Alembertian on a vector potential, and $R{}_{\mu \nu
} $ denotes the Ricci tensor. However, by direct computation using Eqs. (\ref%
{ds1})-(\ref{Hc}), it turns out that%
\begin{equation}
R_{\mu \nu }=\mathcal{O}\left( {h^{2}}\right) .  \label{ri}
\end{equation}%
Therefore, the spatial part of Eq. (\ref{me}) can now be reduced:\footnote[1]{%
The temporal part of the Maxwell's equations can be handled trivially by
fixing the residual gauge within the Lorenz gauge; namely, the radiation
gauge. In this gauge, one can disregard the scalar potential $A^{0}$, as it
becomes zero in charge-free regions (or regions far from electric charge).}
\begin{equation}
\square A^{i}=\mathcal{O}\left( {h^{2}}\right) ,  \label{me1}
\end{equation}%
where $i$ refers to the coordinates $\left( x,y,z\right) $. This can be
regarded as a homogeneous vector wave equation to first order in $h$.

Now, we aim to obtain a decomposition solution for Eq. (\ref{me1}) via
perturbation in $h$:%
\begin{equation}
A^{i}=A_{\mathrm{o}}^{i}+\delta A_{[{h}]}^{i}+\mathcal{O}\left( {h^{2}}%
\right) ,  \label{sol}
\end{equation}%
where $A_{\mathrm{o}}^{i}$ denotes the zeroth-order, unperturbed solution
and $\delta A_{[{h}]}^{i}$ denotes the first-order perturbation solution.
Then one may recast the left-hand side of Eq. (\ref{me1}) as%
\begin{equation}
\square A^{i}=\square _{\mathrm{o}}A_{\mathrm{o}}^{i}+\square _{\mathrm{o}%
}\delta A_{[{h}]}^{i}+\square _{\lbrack {h}]}A_{\mathrm{o}}^{i}+\mathcal{O}%
\left( {h^{2}}\right) ,  \label{lme}
\end{equation}%
where $\square _{\mathrm{o}}\equiv -c^{-2}\partial ^{2}/\partial
t^{2}+\partial ^{2}/\partial x^{2}+\partial ^{2}/\partial y^{2}+\partial
^{2}/\partial z^{2}$ denotes the flat d'Alembertian and $\square _{\lbrack {h%
}]}A_{\mathrm{o}}^{i}$ means the $\mathcal{O}\left( {h}\right) $ piece
remaining from $\square A_{\mathrm{o}}^{i}-\square _{\mathrm{o}}A_{\mathrm{o}%
}^{i}$. Rearranging the terms in Eq. (\ref{lme}) for order-by-order
perturbation, we obtain%
\begin{eqnarray}
\square _{\mathrm{o}}A_{\mathrm{o}}^{i} &=&0~\text{(unperturbed)},
\label{p0} \\
\square _{\mathrm{o}}\delta A_{[{h}]}^{i} &=&-\square _{\lbrack {h}]}A_{%
\mathrm{o}}^{i}~\text{(first order in }h\text{)},  \label{p1}
\end{eqnarray}%
where the first equation implies that $A_{\mathrm{o}}^{i}$ is a solution for
the \textit{unperturbed} homogeneous wave equation defined in flat
spacetime, and the second equation implies that $\delta A_{[{h}]}^{i}$ is a
solution for the \textit{first-order perturbed} inhomogeneous wave equation
defined in flat spacetime with a source term $-\square _{\lbrack {h}]}A_{%
\mathrm{o}}^{i}$, which is also first-order perturbed. It should be noted
here that $\delta A_{[{h}]}^{i}$ can be obtained only after $A_{\mathrm{o}%
}^{i}$ is known: the source term for the first-order perturbed equation
requires knowledge of $A_{\mathrm{o}}^{i}$.

To first order in $h$, the total solution as given from (\ref{sol}) is%
\begin{equation}
A_{\mathrm{total}}^{i}\left( t,\mathbf{x}\right) =A_{\mathrm{o}}^{i}\left( t,%
\mathbf{x}\right) +\delta A_{[{h}]}^{i}\left( t,\mathbf{x}\right) .
\label{sol2}
\end{equation}%
Suppose that the initial unperturbed light is linearly polarized and
propagates along the direction of the wave vector $\mathbf{K}=\left(
K_{x},K_{y},K_{z}\right) $. Then one can write down a solution to satisfy
Eq. (\ref{p0}):%
\begin{align}
A_{\mathrm{o}}^{i}\left( t,\mathbf{x}\right) & =\left( -\frac{K_{y}}{\sqrt{%
K_{x}^{2}+K_{y}^{2}}}\delta _{x}^{i}+\frac{K_{x}}{\sqrt{K_{x}^{2}+K_{y}^{2}}}%
\delta _{y}^{i}\right)  \notag \\
& \times \mathcal{A}\exp \left[ \mathrm{i}\left( \mathbf{K\cdot x}-\omega _{%
\mathrm{e}}t\right) \right] ,  \label{sol3}
\end{align}%
where $\mathcal{A}$ represents the amplitude of EMW, and $\omega _{\mathrm{e}%
}$ denotes the EMW frequency; $\omega _{\mathrm{e}}=cK$ with $c$ being the
speed of light and $K=\sqrt{K_{x}^{2}+K_{y}^{2}+K_{z}^{2}}$ being the
wavenumber for EMW.\footnote[2]{%
Our analysis can be extended to circular and elliptical polarization by
expressing the unperturbed light as $A_{\mathrm{o}}^{i}\left( t,\mathbf{x}%
\right) =\left[ \left( -\frac{K_{y}}{\sqrt{K_{x}^{2}+K_{y}^{2}}}\delta
_{x}^{i}+\frac{K_{x}}{\sqrt{K_{x}^{2}+K_{y}^{2}}}\delta _{y}^{i}\right)
\right. $%
\par
$\left. +\left( -\frac{K_{z}K_{x}}{K\sqrt{K_{x}^{2}+K_{y}^{2}}}\delta
_{x}^{i}-\frac{K_{y}K_{z}}{K\sqrt{K_{x}^{2}+K_{y}^{2}}}\delta _{y}^{i}+\frac{%
\sqrt{K_{x}^{2}+K_{y}^{2}}}{K}\delta _{z}^{i}\right) \exp \left( \mathrm{i}%
\varphi \right) \right] $%
\par
$\times \mathcal{A}\exp \left[ \mathrm{i}\left( \mathbf{K\cdot x}-\omega _{%
\mathrm{e}}t\right) \right] $, where $\varphi $ denotes the relative phase
difference. The light is circularly polarized for $\left\vert \varphi
\right\vert =\frac{\pi }{2}$ and elliptically polarized for $0<\left\vert
\varphi \right\vert <\frac{\pi }{2}$.} Note here that the direction of
polarization is set perpendicular to $\mathbf{K}$. Now, using Eq. (\ref{sol3}%
) for Eq. (\ref{p1}), and by straightforward but tedious computation, we
obtain a perturbation solution $\delta A_{[{h}]}^{i}\left( t,\mathbf{x}%
\right) $.\footnote[3]{%
M\textsc{aple} and \textsc{gr}T\textsc{ensor} have been used extensively to
obtain the results reported here.} To the full, it turns out that $\delta
A_{[{h}]}^{i}\sim \mathcal{O}\left( {h}\right) +\left( \omega _{\mathrm{e}%
}/\omega _{\mathrm{g}}\right) \mathcal{O}\left( {h}\right) $. However,
practically, $\omega _{\mathrm{e}}\gg \omega _{\mathrm{g}}$ (e.g., $\omega _{%
\mathrm{e}}/\omega _{\mathrm{g}}\sim 10^{9}$ to $10^{14}$ for LIGO, $10^{12}$
to $10^{19}$ for LISA, $10^{14}$ to $10^{17}$ for PTA etc.), and therefore
the part $\left( \omega _{\mathrm{e}}/\omega _{\mathrm{g}}\right) \mathcal{O}%
\left( {h}\right) $ would be the only meaningful piece to take for our
analysis; that is, this piece remains in the geometrical-optics
approximation. This finally enables us to express the solution: 
\begin{equation}
\delta A_{[{h}]}^{i}\left( t,\mathbf{x}\right) =2\left( \omega _{\mathrm{e}%
}/\omega _{\mathrm{g}}\right) A_{\mathrm{o}}^{i}\left( t,\mathbf{x}\right) 
\mathcal{H}\left( t,\mathbf{x};\mathbf{K},\mathbf{k}\right) ,  \label{sol4}
\end{equation}%
where 
\begin{eqnarray}
\mathcal{H}\left( t,\mathbf{x};\mathbf{K},\mathbf{k}\right) &\equiv &h_{+}%
\mathcal{F}_{+}\left( \phi ,\theta ,\psi ;\mathbf{K}\right) \cos \left( 
\mathbf{k\cdot x}-\omega _{\mathrm{g}}t\right)  \notag \\
&-&h_{\times }\mathcal{F}_{\times }\left( \phi ,\theta ,\psi ;\mathbf{K}%
\right) \sin \left( \mathbf{k\cdot x}-\omega _{\mathrm{g}}t\right) ,
\label{H}
\end{eqnarray}%
where 
\begin{align}
& \mathcal{F}_{+}\left( \phi ,\theta ,\psi ;\mathbf{K}\right)  \notag \\
& \equiv \frac{1}{2\mathcal{D}}\left( 1+\frac{K_{x}\sin \theta \cos \phi
+K_{y}\sin \theta \sin \phi +K_{z}\cos \theta }{K}\right)  \notag \\
& \times \left\{ \left[ K_{x}^{2}\left( -\cos ^{2}\theta \cos ^{2}\phi +\sin
^{2}\phi \right) \right. \right.  \notag \\
& \left. \left. -K_{x}K_{y}\left( 1+\cos ^{2}\theta \right) \sin \left(
2\phi \right) \right. \right.  \notag \\
& \left. \left. +K_{x}K_{z}\sin \left( 2\theta \right) \cos \phi
+K_{y}^{2}\left( -\cos ^{2}\theta \sin ^{2}\phi +\cos ^{2}\phi \right)
\right. \right.  \notag \\
& \left. \left. +K_{y}K_{z}\sin \left( 2\theta \right) \sin \phi
-K_{z}^{2}\sin ^{2}\theta \right] \cos \left( 2\psi \right) \right.  \notag
\\
& \left. +\left[ K_{x}^{2}\cos \theta \sin \left( 2\phi \right)
-2K_{x}K_{y}\cos \theta \cos \left( 2\phi \right) \right. \right.  \notag \\
& \left. \left. -2K_{x}K_{z}\sin \theta \sin \phi -K_{y}^{2}\cos \theta \sin
\left( 2\phi \right) \right. \right.  \notag \\
& \left. \left. +2K_{y}K_{z}\sin \theta \cos \phi _{_{{}}}^{^{{}}}\right]
\sin \left( 2\psi \right) \right\} ,  \label{C}
\end{align}%
with 
\begin{align}
\mathcal{D}& \equiv K_{x}^{2}\left( 1-\sin ^{2}\theta \cos ^{2}\phi \right)
-K_{x}K_{y}\sin ^{2}\theta \sin \left( 2\phi \right)  \notag \\
& -K_{x}K_{z}\sin \left( 2\theta \right) \cos \phi +K_{y}^{2}\left( 1-\sin
^{2}\theta \sin ^{2}\phi \right)  \notag \\
& -K_{y}K_{z}\sin \left( 2\theta \right) \sin \phi +K_{z}^{2}\sin ^{2}\theta
,  \label{D}
\end{align}%
and%
\begin{equation}
\mathcal{F}_{\times }\left( \phi ,\theta ,\psi ;\mathbf{K}\right) =\mathcal{F%
}_{+}\left( \phi ,\theta ,\psi -\pi /4;\mathbf{K}\right) .  \label{S}
\end{equation}%
Further, the expression in (\ref{C}) can be reduced to a compact form:%
\begin{align}
& \mathcal{F}_{+}\left( \phi ,\theta ,\psi ;\phi _{\star },\theta _{\star
}\right)  \notag \\
& =\frac{\cos ^{2}\gamma _{2}\cos \left( 2\psi \right) -2\cos \gamma
_{2}\sin \theta _{\star }\sin \left( \phi -\phi _{\star }\right) \sin \left(
2\psi \right) }{2\left( 1-\cos \gamma _{1}\right) },  \label{C1}
\end{align}%
where\footnote[4]{%
From (\ref{cg1}) and (\ref{cg2}), one can see that $\gamma _{1}$ is the
angle subtended by an arc between the two points $\left( \theta ,\phi
\right) $ and $\left( \theta _{K},\phi _{K}\right) $ on a unit sphere, while 
$\gamma _{2}$ is the angle subtended by another arc between $\left( \theta
-\pi /2,\phi \right) $ and $\left( \theta _{K},\phi _{K}\right) $ on the
sphere; due to the spherical law of cosines.} 
\begin{eqnarray}
\cos \gamma _{1} &\equiv &\cos \theta \cos \theta _{\star }+\sin \theta \sin
\theta _{\star }\cos \left( \phi -\phi _{\star }\right) ,  \label{cg1} \\
\cos \gamma _{2} &\equiv &\sin \theta \cos \theta _{\star }-\cos \theta \sin
\theta _{\star }\cos \left( \phi -\phi _{\star }\right) ,  \label{cg2}
\end{eqnarray}%
and $\left( \phi _{\star },\theta _{\star }\right) $ have been defined from $%
\left( K_{x},K_{y},K_{z}\right) $ by means of 
\begin{equation}
\sin \theta _{\star }\cos \phi _{\star }=\frac{K_{x}}{K},\sin \theta _{\star
}\sin \phi _{\star }=\frac{K_{y}}{K},\cos \theta _{\star }=\frac{K_{z}}{K}.
\label{Ksph}
\end{equation}%
Also,%
\begin{equation}
\mathcal{F}_{\times }\left( \phi ,\theta ,\psi ;\phi _{\star },\theta
_{\star }\right) =\mathcal{F}_{+}\left( \phi ,\theta ,\psi -\pi /4;\phi
_{\star },\theta _{\star }\right) .  \label{S1}
\end{equation}%
It should be noted again here that the perturbation solution exhibits the
spin-2 tensor modes of the $+$ and $\times $ polarizations through the
dependence on $\psi $ in (\ref{C}) and (\ref{S}) (or (\ref{C1}) and (\ref{S1}%
)).\footnote[5]{%
It can be checked that for $\left( K_{x},K_{y},K_{z}\right) =\left(
0,0,-K\right) $, $\mathcal{F}_{+}\mathcal{\ }$and $\mathcal{F}_{\times }$
reduce to $F_{+}$ and$\ F_{\times }$ in (\ref{ant1}) and (\ref{ant2}),
respectively, which also exhibit the spin-2 tensor modes.}

Eqs. (\ref{sol4})-(\ref{S1}) present the major result of our analysis; it
expresses a perturbation of light due to GWs for a general configuration,
wherein both light and GWs propagate in arbitrary directions. The
perturbation of light will be shown to be equivalent to a delay of the
photon transit time in Sect. \ref{pert}. The equivalence will then be
employed to compute the response\ function for the detection of GWs in Sect. %
\ref{app}. In Appendix \ref{appendA} we show how one can obtain the solution
given by (\ref{sol2})-(\ref{S}) in a computationally tractable manner by
means of coordinate transformations.

\subsection{Perturbed light and delay of photon transit time\label{pert}}

Above we have described how light is perturbed when it propagates in a
spacetime with GWs, by solving Maxwell's equations in that spacetime via a
perturbation method. Suppose that light propagates along the direction of $%
\mathbf{K}=\left( K_{x},K_{y},K_{z}\right) =\left( 0,0,-K\right) $, as in
the example of a PTA to be discussed in Sect. \ref{app}. As $K_{z}=-K<0$,
our light propagates along $-z$ direction; i.e., from the sky towards the
earth. Then it can be expressed by the electric field $E_{\mathrm{total}%
}^{i}\left( t,0,0,z\right) =-c^{-1}\left( \partial /\partial t\right) A_{%
\mathrm{total}}^{i}\left( t,0,0,z\right) $, obtained from Eqs. (\ref{sol2})-(%
\ref{sol4}). Starting at $\left( t,z\right) =\left( t_{0},L\right) $, the
propagation path can be written as $z=L-c\left( t-t_{0}\right) $ for $%
t_{0}\leq t\leq t_{0}+T$, with $L=cT$. Then we find 
\begin{align}
& \left. \frac{\delta E_{\left[ h\right] }^{i}}{E_{\mathrm{o}}^{i}}%
\right\vert _{z=0}-\left. \frac{\delta E_{\left[ h\right] }^{i}}{E_{\mathrm{o%
}}^{i}}\right\vert _{z=L}  \notag \\
& =\frac{\omega _{\mathrm{e}}\left( h_{+}F_{+}+\mathrm{i}h_{\times
}F_{\times }\right) \left\{ 1-\exp \left[ \mathrm{i}kL\left( 1+\cos \theta
\right) \right] \right\} }{\omega _{\mathrm{g}}}  \notag \\
& \times \exp \left[ -\mathrm{i}\left( kL+\omega _{\mathrm{g}}t_{0}\right) %
\right] ,  \label{pl}
\end{align}%
where $E_{\mathrm{o}}^{i}=-c^{-1}\left( \partial /\partial t\right) A_{%
\mathrm{o}}^{i}$, $\delta E_{\left[ h\right] }^{i}=-c^{-1}\left( \partial
/\partial t\right) \delta A_{[{h}]}^{i}$, and the right-hand side is
expressed in the complex representation for analytical convenience, and 
\begin{eqnarray}
F_{+} &\equiv &\sin ^{2}\left( \theta /2\right) \cos \left( 2\psi \right) ,
\label{ant1} \\
F_{\times } &\equiv &\sin ^{2}\left( \theta /2\right) \sin \left( 2\psi
\right) ,  \label{ant2}
\end{eqnarray}%
are antenna patterns for $+$ and $\times $ polarization states, respectively.\footnote[6]{%
Our expressions of antenna patterns are in agreement with those for pulsar
timing arrays in Refs. \cite{Chamberlin,Yunes}.}

On the other hand, when a photon propagates in a spacetime with GWs, its
trajectory will be perturbed, resulting in a delay of its transit time. The
propagation takes place along the null geodesic, i.e., $ds^{2}=0$ in Eq. (%
\ref{ds1}), and hence one can express a delay for a photon propagating by a
distance $L=cT$ along $-z$ direction, starting at $\left( t,z\right) =\left(
t_{0},L\right) $; that is, along the path $z=L-c\left( t-t_{0}\right) $ for $%
t_{0}\leq t\leq t_{0}+T$ \cite{Rakhmanov}:%
\begin{align}
& \frac{\delta T_{\left[ h\right] }}{T}=\frac{1}{2cT}\int_{L}^{0}h_{zz}%
\left( t_{0},0,0,z\right) dz+\mathcal{O}\left( h^{2}\right)  \notag \\
& =-\mathrm{i}\frac{\left( h_{+}F_{+}+\mathrm{i}h_{\times }F_{\times
}\right) \left\{ 1-\exp \left[ \mathrm{i}kL\left( 1+\cos \theta \right) %
\right] \right\} }{kL}  \notag \\
& \times \exp \left[ -\mathrm{i}\left( kL+\omega _{\mathrm{g}}t_{0}\right) %
\right] +\mathcal{O}\left( h^{2}\right) ,  \label{dpt}
\end{align}%
where $\delta T_{\left[ h\right] }$ means the deviation of the transit time
from $T$, and $h_{zz}$ is read off from Eq. (\ref{ds1}) and expressed in the
complex representation for analytical convenience.

Comparing Eqs. (\ref{pl}) and (\ref{dpt}), we establish a relation between
the delay of the photon transit time and the perturbation of light due to
GWs:

\begin{equation}
{{\frac{\delta T_{\left[ {h}\right] }}{T}\simeq }\,\mathcal{N}}\left( \left. 
\frac{\delta E_{\left[ h\right] }^{i}}{E_{\mathrm{o}}^{i}}\right\vert
_{z=0}-\left. \frac{\delta E_{\left[ h\right] }^{i}}{E_{\mathrm{o}}^{i}}%
\right\vert _{z=L}\right) ,  \label{rel}
\end{equation}%
where $\mathcal{N}=\left( \mathrm{i}\omega _{\mathrm{e}}T\right)
^{-1}=\left( \mathrm{i}KL\right) ^{-1}$. Here one can give a physical
interpretation of this relation: light perturbed by GWs, being described by
Maxwell's equations (\ref{me}), leads to a delay of the photon transit time,
being described by the null geodesic equation, $ds^{2}=0$ in (\ref{ds1}).

The relation given by Eq. (\ref{rel}) must be true for a general
configuration in which both light and GWs are assumed to propagate in
arbitrary directions. That is, a delay of the photon transit time along an
arbitrary path can equivalently be computed, using the solutions of
Maxwell's equations for the general case as given by Eqs. (\ref{sol2})-(\ref%
{sol4}). Then the relation (\ref{rel}) is now extended to 
\begin{equation}
{{\frac{\delta T_{\left[ {h}\right] }}{T}\simeq }\,\mathcal{N}}\left( \left. 
\frac{\delta E_{\left[ h\right] }^{i}}{E_{\mathrm{o}}^{i}}\right\vert _{%
\text{earth}}-\left. \frac{\delta E_{\left[ h\right] }^{i}}{E_{\mathrm{o}%
}^{i}}\right\vert _{\text{sky}}\right) ,  \label{rel1}
\end{equation}%
where one can set the `earth' location to be $\left( x,y,z\right) =\left(
0,0,0\right) $ and the `sky' location to be $\left( x,y,z\right) =\left(
-L\sin \theta _{\star }\cos \phi _{\star },-L\sin \theta _{\star }\sin \phi
_{\star },-L\cos \theta _{\star }\right) $, with $\pi \leq \theta _{\star
}\leq 3\pi /2$, for computational convenience. From this and with the
electric field $E_{\mathrm{total}}^{i}\left( t,x,y,z\right) =-c^{-1}\left(
\partial /\partial t\right) A_{\mathrm{total}}^{i}\left( t,x,y,z\right) $
for a general $\mathbf{K}=\left( K_{x},K_{y},K_{z}\right) =\left( K\sin
\theta _{\star }\cos \phi _{\star },K\sin \theta _{\star }\sin \phi _{\star
},K\cos \theta _{\star }\right) $, obtained from Eqs. (\ref{sol2})-(\ref%
{sol4}), the delay of the photon transit time is finally written as 
\begin{align}
& \frac{\delta T_{\left[ h\right] }}{T}  \notag \\
& =-\mathrm{i}\frac{\left( h_{+}\mathcal{F}_{+}+\mathrm{i}h_{\times }%
\mathcal{F}_{\times }\right) \left\{ 1-\exp \left[ \mathrm{i}kL\left( 1-\cos
\gamma _{1}\right) \right] \right\} }{kL}  \notag \\
& \times \exp \left[ -\mathrm{i}\left( kL+\omega _{\mathrm{g}}t_{0}\right) %
\right] +\mathcal{O}\left( h^{2}\right) ,  \label{dpt1}
\end{align}%
where $\mathcal{F}_{+}$, $\mathcal{F}_{\times }$ and $\cos \gamma _{1}$
refer to (\ref{C1}), (\ref{S1}) and (\ref{cg1}), respectively.

\subsection{Application - pulsar timing array (PTA)\label{app}}

\begin{figure}
    \centering
    \resizebox{\columnwidth}{!}{
        \begin{tikzpicture}
            \pic {emgw};
            
            \node at (0,-3.5) {
                \tdplotsetmaincoords{0}{0}
                \begin{tikzpicture}[tdplot_main_coords]
                    \tdplotsetrotatedcoords{30}{100}{200}
                    \draw[-stealth,tdplot_rotated_coords] (0,0,0) -- (3,0,0) node[pos=1.1]{\( y \)};
                    \draw[-stealth,tdplot_rotated_coords] (0,0,0) -- (0,1.5,0) node[pos=1.1]{\( x \)};
                    \draw[-stealth,tdplot_rotated_coords] (0,0,0) -- (0,0,1.35) node[pos=1.1]{\( z \)};
                \end{tikzpicture}
            };
            \node at (0.2,-5) {Detector frame};
            \node at (4.3,-3.7) {
                \tdplotsetmaincoords{0}{0}
                \begin{tikzpicture}[tdplot_main_coords]
                    \tdplotsetrotatedcoords{50}{80}{20}
                    \draw[-stealth,tdplot_rotated_coords] (0,0,0) -- (3,0,0) node[pos=1.3]{\( y' \)};
                    \draw[-stealth,tdplot_rotated_coords] (0,0,0) -- (0,1.5,0) node[pos=1.3]{\( z' \)};
                    \draw[-stealth,tdplot_rotated_coords] (0,0,0) -- (0,0,1.2) node[pos=1.3]{\( x' \)};
                \end{tikzpicture}
            };
            \node at (4.5,-5) {GW frame};
        \end{tikzpicture}
    }
    \caption{An illustration of a pulsar timing array (PTA) together with the detector frame and the GW frame.}
    \label{fig0}
\end{figure}
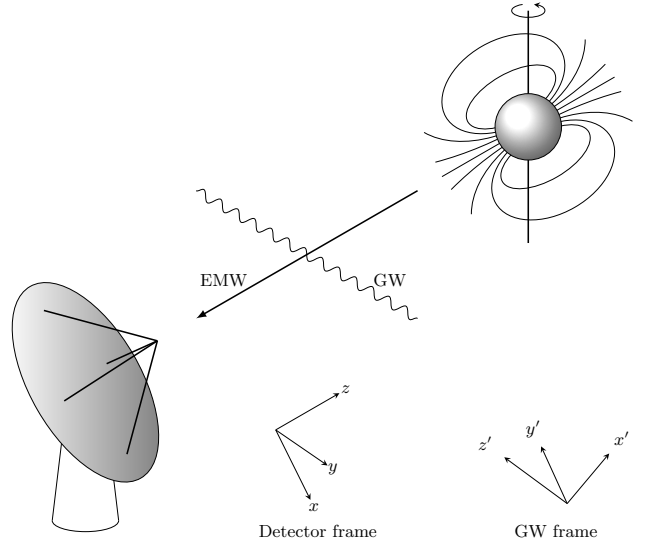

The property of the perturbed light as given by Eq. (\ref{rel1}) (or (\ref%
{rel})) can be applied to the detection of GWs, and for its simplest
application, we consider a PTA. One can arrange a detector (e.g., a radio
telescope) to receive photons emitted from a pulsar to measure pulse arrival
time as illustrated in Fig.~\ref{fig0}. A pulsar can serve as an astronomical clock of excellent precision,
with the constancy of the measured pulse frequency $\nu _{\mathrm{o}}$.
However, with GWs passing through our space, the measured frequency $\nu
\left( t\right) $ will vary slightly. Then the effects of GWs can be
determined from the variation of the frequency (or from the variation of the
pulse period) $\left[ \nu _{\mathrm{o}}-\nu \left( t\right) \right] /\nu _{%
\mathrm{o}}\simeq \left[ \tau \left( t\right) -\tau _{\mathrm{o}}\right]
/\tau _{\mathrm{o}}$, where $\tau \left( t\right) =\nu ^{-1}\left( t\right) $
is the measured pulse period and $\tau _{\mathrm{o}}=\nu _{\mathrm{o}}^{-1}$
is the constancy of the measured pulse period \cite{Sazhin,Detweiler}.

For the cumulative variation, we define a \textquotedblleft
residual\textquotedblright\ \cite{Detweiler}, which can be expressed using
Eq. (\ref{dpt1}) with $T=\tau _{\mathrm{o}}=L/c$ as%
\begin{eqnarray}
\mathfrak{r}\left( t\right) &\equiv &\int_{0}^{t}\frac{\nu _{\mathrm{o}}-\nu
\left( t^{\prime }\right) }{\nu _{\mathrm{o}}}dt^{\prime }\;\simeq
\int_{0}^{t}\frac{\tau \left( t^{\prime }\right) -\tau _{\mathrm{o}}}{\tau _{%
\mathrm{o}}}dt^{\prime }  \notag \\
&\sim &\frac{h_{+}\mathcal{G}_{+}+\mathrm{i}h_{\times }\mathcal{G}_{\times }%
}{f}\exp \left( -2\mathrm{i}\pi ft\right) ,\;  \label{res}
\end{eqnarray}%
where $t^{\prime }\leftarrow t_{0}$ from Eq. (\ref{dpt1}),\ and $f=\omega _{%
\mathrm{g}}/\left( 2\pi \right) $, and%
\begin{align}
\mathcal{G}_{+}& \equiv & & \frac{\mathcal{F}_{+}\exp \left( -\mathrm{i}%
kL\right) \left\{ 1-\exp \left[ 2\mathrm{i}\pi f\tau _{\mathrm{o}}\left(
1-\cos \gamma _{1}\right) \right] \right\} }{4\pi ^{2}f\tau _{\mathrm{o}}},
\label{ant3} \\
\mathcal{G}_{\times }& \equiv & & \frac{\mathcal{F}_{\times }\exp \left( -%
\mathrm{i}kL\right) \left\{ 1-\exp \left[ 2\mathrm{i}\pi f\tau _{\mathrm{o}%
}\left( 1-\cos \gamma _{1}\right) \right] \right\} }{4\pi ^{2}f\tau _{%
\mathrm{o}}},  \label{ant4}
\end{align}%
with $\mathcal{F}_{+}$, $\mathcal{F}_{\times }$ and $\cos \gamma _{1}$ given
by (\ref{C1}), (\ref{S1}) and (\ref{cg1}), respectively. Here $\mathcal{G}%
_{+}$ and $\mathcal{G}_{\times }$ are termed \textit{exact} detector
responses, and for $\mathbf{K}=\left( K_{x},K_{y},K_{z}\right) =\left(
0,0,-K\right) $ in particular, they reduce to 
\begin{align}
G_{+}& \equiv \frac{F_{+}\exp \left( -\mathrm{i}kL\right) \left\{ 1-\exp %
\left[ 2\mathrm{i}\pi f\tau _{\mathrm{o}}\left( 1+\cos \theta \right) \right]
\right\} }{4\pi ^{2}f\tau _{\mathrm{o}}},  \label{ant5} \\
G_{\times }& \equiv \frac{F_{\times }\exp \left( -\mathrm{i}kL\right)
\left\{ 1-\exp \left[ 2\mathrm{i}\pi f\tau _{\mathrm{o}}\left( 1+\cos \theta
\right) \right] \right\} }{4\pi ^{2}f\tau _{\mathrm{o}}},  \label{ant6}
\end{align}%
respectively, with $F_{+}$ and $F_{\times }$ given by (\ref{ant1}) and (\ref%
{ant2}). 
\begin{figure}[tbp]
\caption{Antenna patterns of the detector responses: (A) $\left\vert
G_{+,\times }\right\vert $ at $f\ll 1\,\mathrm{Hz}$, (B) $\left\vert
G_{+,\times }\right\vert $ at $f=1000\,\mathrm{Hz}$ for a\ light ray with $%
\left( K_{x},K_{y},K_{z}\right) =\left( 0,0,-K\right) $ from a millisecond
pulsar with $\protect\tau _{\mathrm{o}}\sim 10^{-3}\,\mathrm{s}$; (C) $%
\left\vert \mathcal{G}_{+,\times }\right\vert $ at $\protect\theta _{\star }=%
\protect\pi +\cos ^{-1}\left( \protect\sqrt{15/23}\right) $, (D) $\left\vert 
\mathcal{G}_{+,\times }\right\vert $ at $\protect\theta _{\star }=3\protect%
\pi /2$ for light rays with $\left( K_{x},K_{y},K_{z}\right) =\left( K\sin 
\protect\theta _{\star }\cos \protect\phi _{\star },K\sin \protect\theta %
_{\star }\sin \protect\phi _{\star },K\cos \protect\theta _{\star }\right) $
from millisecond pulsars with $\protect\tau _{\mathrm{o}}\sim 10^{-3}\,%
\mathrm{s}$, both being considered in the low-frequency regime, $f\ll 1\,%
\mathrm{Hz}$. Here the 3D plots are drawn in (A) and (B) from $\left\vert G_{+,\times
}\left( \protect\theta ,\protect\psi \right) \right\vert $, and in (C) and (D) 
from $\left\vert \mathcal{G}_{+,\times }\left( \protect\theta ,\protect\psi %
\right) \right\vert \equiv \left\vert \protect\int_{0}^{2\protect\pi }%
\mathcal{G}_{+,\times }\left( \protect\phi ,\protect\theta ,\protect\psi %
\right) d\protect\phi /2\protect\pi \right\vert $, in a
Cartesian coordinate frame via $\left( X,Y,Z\right) =\left( \sin \protect%
\theta \cos \protect\psi ,\sin \protect\theta \sin \protect\psi ,\cos 
\protect\theta \right) $. Note that the patterns for $\left\vert \mathcal{G}%
_{+}\right\vert $ and $\left\vert \mathcal{G}_{\times }\right\vert $ (also $%
\left\vert G_{+}\right\vert $ and $\left\vert G_{\times }\right\vert $) are
identical except for the rotational phase difference $\protect\pi /4$, due
to $\mathcal{G}_{+}\left( \protect\theta ,\protect\psi \right) =\mathcal{G}%
_{\times }\left( \protect\theta ,\protect\psi +\protect\pi /4\right) $ (also 
$G_{+}\left( \protect\theta ,\protect\psi \right) =G_{\times }\left( \protect%
\theta ,\protect\psi +\protect\pi /4\right) $) from Eqs. (\protect\ref{ant3}%
) and (\protect\ref{ant4}) (also Eqs. (\protect\ref{ant5}) and (\protect\ref%
{ant6})).}
\label{fig1}\includegraphics[width=8.6cm]{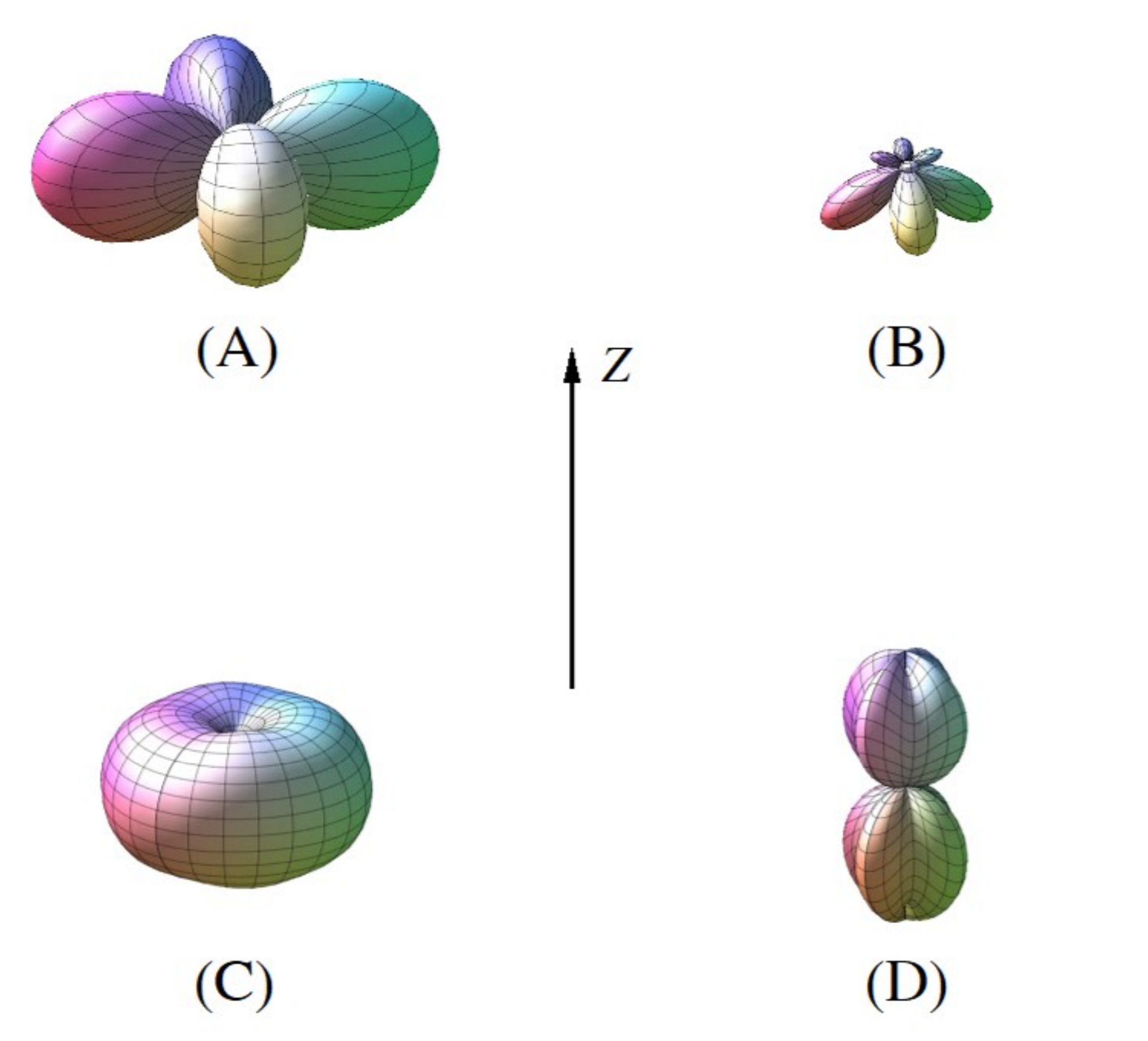}
\end{figure}

Out of Eq. (\ref{res}), one can express 
\begin{align}
& \left\langle \mathfrak{r}^{2}\left( t\right) \right\rangle _{\mathrm{time}%
}\;\sim \;f^{2}\tilde{\mathfrak{r}}\left( f\right) \tilde{\mathfrak{r}}%
^{\ast }\left( f\right)  \notag \\
& \simeq \left\vert \mathcal{G}_{+}\left( f\right) \right\vert
^{2}\left\vert \tilde{h}_{+}\left( f\right) \right\vert ^{2}+\left\vert 
\mathcal{G}_{\times }\left( f\right) \right\vert ^{2}\left\vert \tilde{h}%
_{\times }\left( f\right) \right\vert ^{2},  \label{avg}
\end{align}%
where $\tilde{\mathfrak{r}}\left( f\right) $, $\tilde{h}_{+}\left( f\right) $
and $\tilde{h}_{\times }\left( f\right) $ denote the Fourier transforms of $%
\mathfrak{r}\left( t\right) $, $h_{+}\left( t\right) \equiv h_{+}\exp \left(
-2\mathrm{i}\pi ft\right) $ and $h_{\times }\left( t\right) \equiv h_{\times
}\exp \left( -2\mathrm{i}\pi ft\right) $, respectively, and $\ast $ denotes
the complex conjugate.

The detector response\ function can be computed by taking a \textit{sky
average} of $\mathcal{G}_{+}\mathcal{G}_{+}^{\ast }+\mathcal{G}_{\times }%
\mathcal{G}_{\times }^{\ast }$ over $\left( \phi ,\theta ,\psi \right) $,
with $\mathcal{G}_{+}$ and $\mathcal{G}_{\times }$ given by (\ref{ant3}) and
(\ref{ant4}), respectively:%
\begin{eqnarray}
\mathcal{R}\left( f\right) \; &\equiv &\;\frac{1}{4\pi ^{2}}\int_{0}^{\pi
}d\psi \int_{0}^{2\pi }d\phi \int_{0}^{\pi }d\theta \,\sin \theta   \notag \\
&&\left[ \mathcal{G}_{+}\left( f\right) \mathcal{G}_{+}^{\ast }\left(
f\right) +\mathcal{G}_{\times }\left( f\right) \mathcal{G}_{\times }^{\ast
}\left( f\right) \right] .  \label{rep}
\end{eqnarray}%
For the special case with $\mathbf{K}=\left( K_{x},K_{y},K_{z}\right)
=\left( 0,0,-K\right) $, the detector responses reduce, that is, $\mathcal{G}%
_{+}\rightarrow G_{+}$ and $\mathcal{G}_{\times }\rightarrow G_{\times }$,
and we obtain%
\begin{equation}
\mathcal{R}\left( f\right) =\frac{32\pi ^{3}f^{3}\tau _{\mathrm{o}%
}^{3}-12\pi f\tau _{\mathrm{o}}+3\sin \left( 4\pi f\tau _{\mathrm{o}}\right) 
}{768\pi ^{7}f^{5}\tau _{\mathrm{o}}^{5}},  \label{rep0}
\end{equation}%
which is a complete closed-form expression. In the limit $f\tau _{\mathrm{o}%
}\ll 1$, which is appropriate for ultra-low-frequency GW signals carried via
millisecond pulsars, this can be approximated as 
\begin{equation}
\mathcal{R}\approx \frac{1}{30\pi ^{2}}+\mathcal{O}\left( f^{2}\tau _{%
\mathrm{o}}^{2}\right) .  \label{Rappr0}
\end{equation}%
However, for a general case with $\mathbf{K}=\left( K_{x},K_{y},K_{z}\right)
=\left( K\sin \theta _{\star }\cos \phi _{\star },K\sin \theta _{\star }\sin
\phi _{\star },K\cos \theta _{\star }\right) $, the computation of the
response\ function is rather involved, and it cannot be computed fully
symbolically unlike (\ref{rep0}) for the special case; but its approximation
can be obtained in the limit $f\tau _{\mathrm{o}}\ll 1$ instead: 
\begin{equation}
\mathcal{R}\approx \frac{29+150\cos ^{2}\theta _{\star }-115\cos ^{4}\theta
_{\star }}{1920\pi ^{2}}+\mathcal{O}\left( f^{2}\tau _{\mathrm{o}%
}^{2}\right) .  \label{Rappr}
\end{equation}%
This manifests how the detector response function varies with $\theta
_{\star }$, the incident angle of a light pulse with respect to the detector.

\begin{figure}[tbp]
\includegraphics[width=8.6cm]{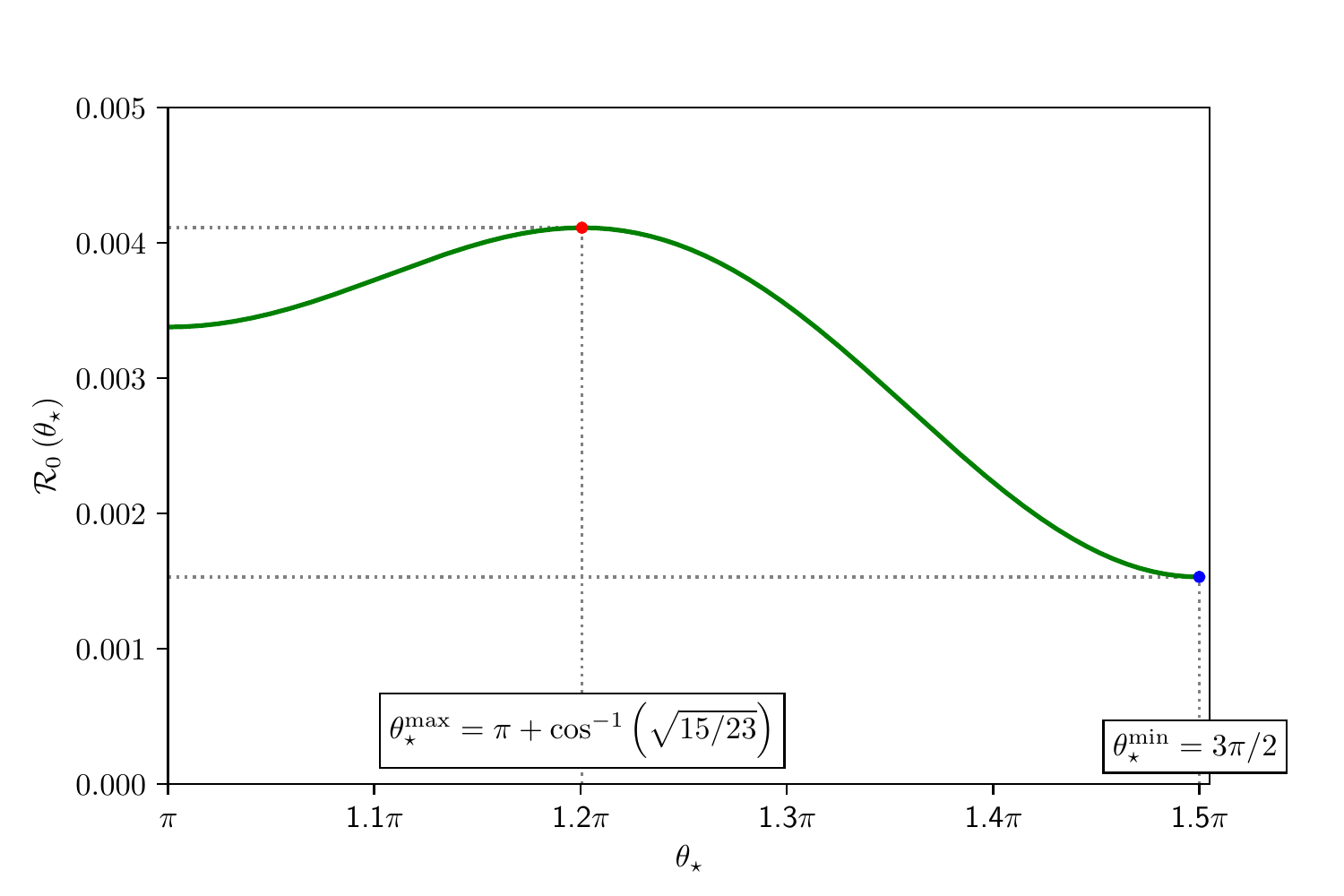}
\caption{A plot of $\mathcal{R}_{0}\left( \protect\theta _{\star }\right)
=\left( 29+150\cos ^{2}\protect\theta _{\star }-115\cos ^{4}\protect\theta %
_{\star }\right) /\left( 1920\protect\pi ^{2}\right) $, the leading order
term in (\protect\ref{Rappr}). Note that $\mathcal{R}_{0}$ becomes maximum
and minimum at $\protect\theta _{\star }=\protect\pi +\cos ^{-1}\left( 
\protect\sqrt{15/23}\right) $ and $3\protect\pi /2$, respectively.}
\label{fig2}
\end{figure}

For a\ light ray with $\left( K_{x},K_{y},K_{z}\right) =\left( 0,0,-K\right) 
$ from a millisecond pulsar with $\tau _{\mathrm{o}}\sim 10^{-3}\,\mathrm{s}$%
, the antenna patterns $\left\vert G_{+,\times }\right\vert $ of the
detector responses (\ref{ant5}) and (\ref{ant6}) are illustrated at
different frequencies, (A) $f\ll 1\,\mathrm{Hz}$ and (B) $f=1000\,\mathrm{Hz}
$, in the upper panel of Fig. \ref{fig1}. Note the decrease in the volume of
the plot for the higher frequency. On the other hand, in the lower panel of
Fig. \ref{fig1}, we illustrate the antenna patterns $\left\vert \mathcal{G}%
_{+,\times }\right\vert $ of the detector responses (\ref{ant3}) and (\ref%
{ant4}) for light rays with $\left( K_{x},K_{y},K_{z}\right) =\left( K\sin
\theta _{\star }\cos \phi _{\star },K\sin \theta _{\star }\sin \phi _{\star
},K\cos \theta _{\star }\right) $ from millisecond pulsars with $\tau _{%
\mathrm{o}}\sim 10^{-3}\,\mathrm{s}$, incident on a detector at different
polar angles, (C) $\theta _{\star }=\pi +\cos ^{-1}\left( \sqrt{15/23}%
\right) $ and (B) $\theta _{\star }=3\pi /2$, which correspond to the
maximum and the minimum detector responses, respectively as determined from (%
\ref{Rappr}), both being considered in the low-frequency regime, $f\ll 1\,%
\mathrm{Hz}$ (see Fig. \ref{fig2}). Note the difference in the volume
between the two plots.

\begin{figure}[tbp]
\includegraphics[width=8.6cm]{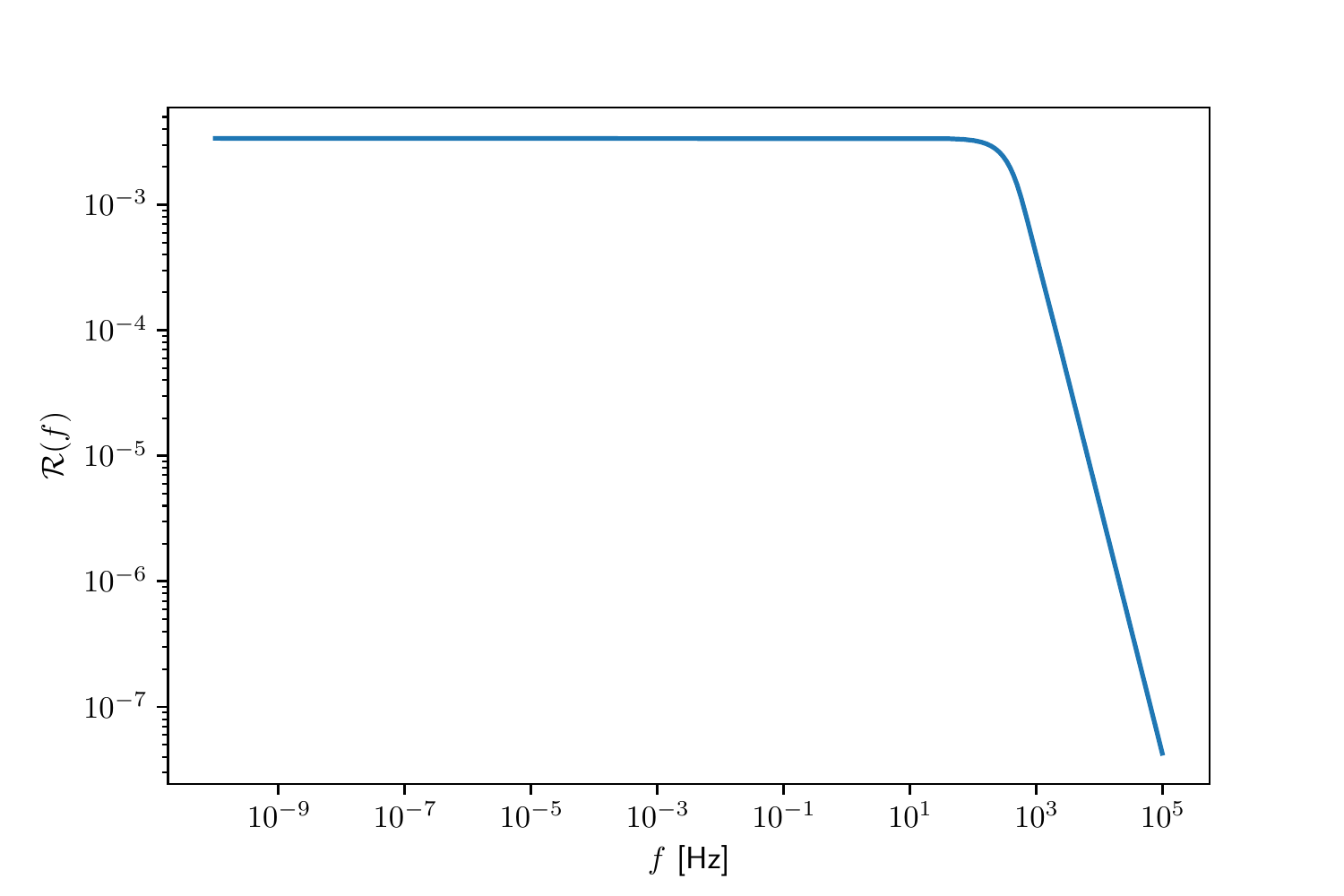}
\caption{A plot of $\mathcal{R}\left( f\right) $ for the special case with $%
\left( K_{x},K_{y},K_{z}\right) =\left( 0,0,-K\right) $ for a millisecond
pulsar with $\protect\tau _{\mathrm{o}}\sim 10^{-3}\,\mathrm{s}$. }
\label{fig3}
\end{figure}

In Fig. \ref{fig2} is plotted $\mathcal{R}_{0}$, the leading-order
approximation of $\mathcal{R}$ in the limit $f\tau _{\mathrm{o}}\ll 1$ in (%
\ref{Rappr}), as a function of the incident angle $\theta _{\star }$ of a
light ray. It should be noted here that $\mathcal{R}_{0}$ becomes maximum at 
$\theta _{\star }=\pi +\cos ^{-1}\left( \sqrt{15/23}\right) $ and minimum at 
$\theta _{\star }=3\pi /2$; that is, no extrema at $\theta _{\star }=\pi $,
i.e., for the special case with $\left( K_{x},K_{y},K_{z}\right) =\left(
0,0,-K\right) $. In Fig. \ref{fig3}\ is shown a plot of $\mathcal{R}\left(
f\right) $ for the special case with $\left( K_{x},K_{y},K_{z}\right)
=\left( 0,0,-K\right) $ for a millisecond pulsar with $\tau _{\mathrm{o}%
}\sim 10^{-3}\,\mathrm{s}$. 

\begin{figure}[tbp]
\includegraphics[width=8.6cm]{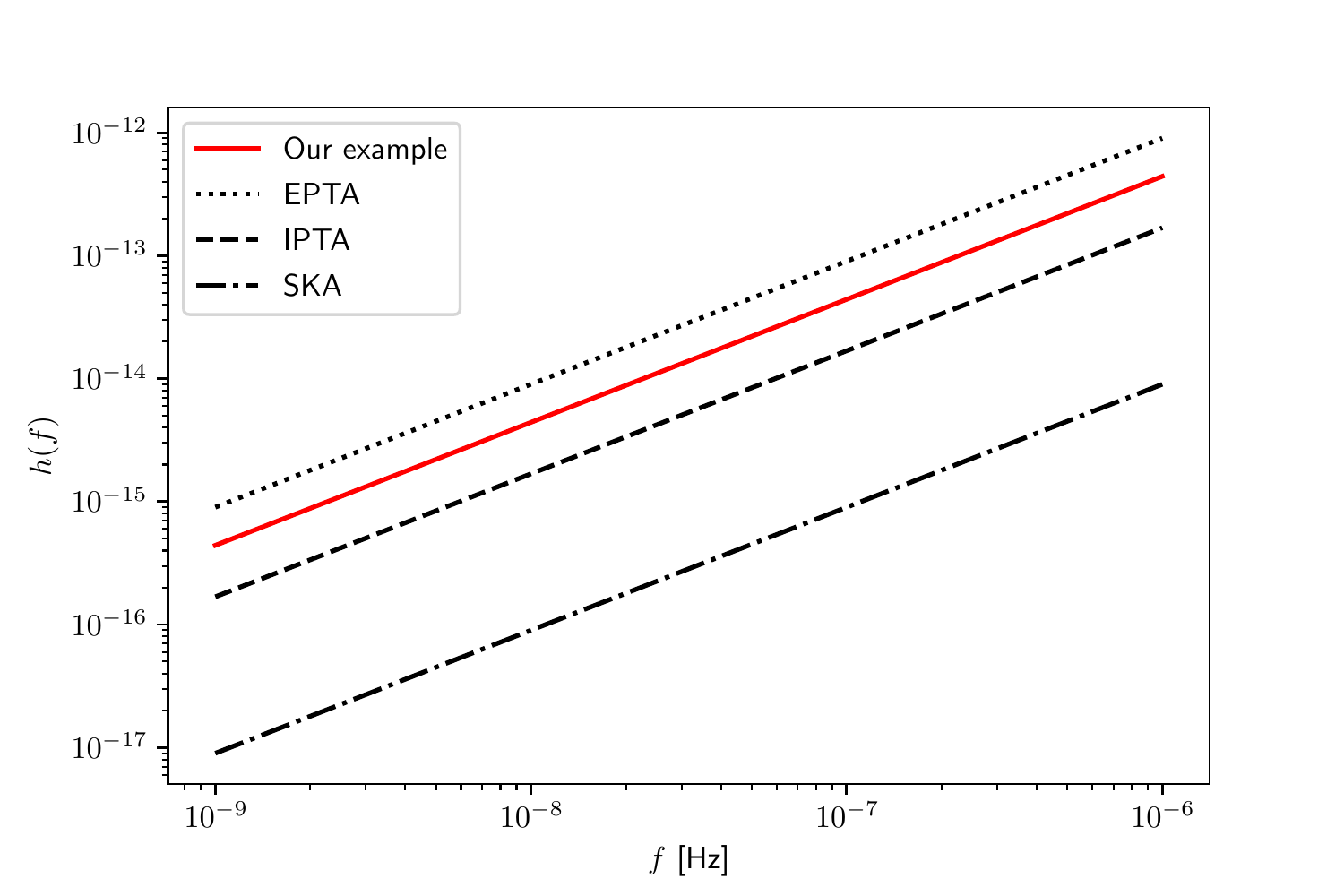}
\caption{A plot of $h\left( f\right) $ (red solid line) for GWs from a
source with $M\sim 10^{9}M_{\odot }$, $R_{\mathrm{o}}\sim 2\times
10^{11}M_{\odot }$ and $r$ $\sim 10^{10}\mathrm{\mathrm{ly}}$, detected by
means of a millisecond pulsar with $\protect\tau _{\mathrm{o}}\sim 10^{-3}\,%
\mathrm{s}$. It is compared with the sensitivity curves for EPTA (black
dotted line), IPTA (black dashed line) and SKA (black dash-dotted line)
(taken from Ref. \protect\cite{Moore}). }
\label{fig4}
\end{figure}

In view of Eqs. (\ref{avg}) and (\ref{rep}), one can determine the detector
\textquotedblleft sensitivity\textquotedblright : 
\begin{equation}
h\left( f\right) \;\equiv \;f\tilde{h}\left( f\right) \;\;\sim \;\sqrt{\frac{%
f^{2}\left\langle \mathfrak{r}^{2}\left( t\right) \right\rangle _{\mathrm{%
time}}}{\mathcal{R}\left( f\right) }}.  \label{sen}
\end{equation}%
Now, following Ref. \cite{Detweiler}, for a periodic GW source, we consider
two supermassive black holes of mass $M$ in a circular orbit of radius $R_{%
\mathrm{o}}$, with the distance $r$ from us. Then one can estimate

\begin{equation}
\sqrt{\left\langle \mathfrak{r}^{2}\left( t\right) \right\rangle }\sim
\omega _{\mathrm{g}}^{-1}h_{\mathrm{max}},  \label{rms}
\end{equation}%
with the maximum strain amplitude and the GW frequency being estimated
respectively as%
\begin{equation}
h_{\mathrm{max}}\;\sim \;5\times 10^{-14}\left( \frac{200M}{R_{\mathrm{o}}}%
\right) \left( \frac{M}{10^{10}M_{\odot }}\right) \left( \frac{10^{10}%
\mathrm{\mathrm{ly}}}{r}\right) ,  \label{hmax}
\end{equation}%
\begin{equation}
\omega _{\mathrm{g}}\;\sim \;2\times 10^{-8}\mathrm{s}^{-1}\left( \frac{200M%
}{R_{\mathrm{o}}}\right) ^{3/2}\left( \frac{10^{10}M_{\odot }}{M}\right) .
\label{GWf}
\end{equation}%
With consideration of multiple pulsars as carriers of the GW signals, one
can use the detector response function averaged over $\pi \leq \theta
_{\star }\leq 3\pi /2$ from (\ref{Rappr}), to evaluate $h\left( f\right) $
via (\ref{sen}):%
\begin{equation}
\mathcal{\bar{R}}\approx \frac{487}{15360\pi ^{2}}+\mathcal{O}\left(
f^{2}\tau _{\mathrm{o}}^{2}\right) ,  \label{Ravg}
\end{equation}%
where the term $\mathcal{O}\left( f^{2}\tau _{\mathrm{o}}^{2}\right) $ can
be disregarded for low-frequency GWs with $f\sim 10^{-8}\,\mathrm{Hz}$ to be
detected via millisecond pulsars with $\tau _{\mathrm{o}}\sim 10^{-3}\,%
\mathrm{s}$. Then, for example, for GWs from a source with $M\sim
10^{9}M_{\odot }$, $R_{\mathrm{o}}\sim 2\times 10^{11}M_{\odot }$ and $r$ $%
\sim 10^{10}\mathrm{\mathrm{ly}}$, we obtain a curve for $h\left( f\right) $
using Eqs. (\ref{sen})-(\ref{Ravg}), as given by Fig. \ref{fig4}; it
compares well with the actual sensitivity curves for EPTA, IPTA and SKA in
the literature \cite{Moore}.

\section{Summary and Discussion\label{disc}}

From a general relativistic perspective, the interaction of light with GWs
can be viewed as equivalent to a perturbation of light due to GWs. We have
solved Maxwell's equations in a spacetime perturbed by GWs and obtained a
solution for a general case as given by Eqs. (\ref{sol2})-(\ref{sol4}),
wherein both light and GWs are assumed to propagate in arbitrary directions.
Based on this solution, it has been shown that a perturbation of light due
to GWs leads to a delay of the photon transit time, as given by Eq. (\ref%
{rel1}). Applying this principle to a PTA, we have worked out the detector
response function $\mathcal{R}$ as given by Eq. (\ref{Rappr}) and Fig. \ref%
{fig2}, which manifests how the detector response varies with the incident
angle of a light pulse with respect to the detector. Then using this, we
have obtained the curve for $h\left( f\right) $ as given by Fig. \ref{fig4}.
Our result shows good agreement with the literature; the $h\left( f\right) $
curve compares well with the actual sensitivity curves for EPTA, IPTA and
SKA, taken from Ref. \cite{Moore}. However, our purpose in this analysis is
rather to check how properly our detector response function $\mathcal{R}$
serves to provide the $h\left( f\right) $ curve for a given GW signal in the
desired order of magnitude. A practical analysis of the detection
sensitivity for the actual PTAs would be based on the accurate measurements
of the timing residuals from multiple pulsars, with the consideration of
systematics in the residuals such as solar system ephemeris errors mimicking
a GW signal, the solar system metric contributing with an extra time delay
in the modeled signal, etc.

As a means of improving our analysis in relation to a PTA, it is worth
considering the cross-correlation of the residuals of two pulsars nearby in
the sky \cite{Detweiler}: the statistically-significant quadrupolar
interpulsar correlation of GW background-induced timing delays as addressed
in Ref. \cite{Hellings}. This will be discussed further in follow-up studies.

Our analysis can be extended to more complex arrays for GW detection than a
PTA. For interferometers such as LIGO and LISA, we require a description of
light rays in more complicated configurations, based on Eqs. (\ref{sol2})-(%
\ref{sol4}). We leave further discussion of this to follow-up studies.

\begin{acknowledgments}
D.-H. Kim was supported by the Basic Science Research Program through the National Research Foundation of Korea (NRF) funded by the Ministry of Education (NRF-2018R1D1A1B07051276 and NRF-2021R1I1A1A01054781). C. Park was supported in part by the Basic Science Research Program through the National Research Foundation of Korea (NRF) funded by the Ministry of Education (NRF-2018R1D1A1B07041004), and by the National Institute for Mathematical Sciences (NIMS) funded by Ministry of Science and ICT (B20710000).
\end{acknowledgments}

\appendix

\section{Solutions to Maxwell's equations via coordinate transformations 
\label{appendA}}

The total decomposition solution (\ref{sol2}) can be obtained in the easiest
manner for a particular case, in which $A_{\mathrm{o}}^{i}$ takes the
simplest (but not trivial) form and so does $\delta A_{[{h}]}^{i}$ as
obtained from Eq. (\ref{p1}). In a particular frame of the coordinates $%
\mathbf{x}^{\prime \prime }\equiv \left( x^{\prime \prime },y^{\prime \prime
},z^{\prime \prime }\right) $, one can prescribe the simplest solution to
satisfy Eq. (\ref{p0}): 
\begin{equation}
A_{\mathrm{o}}^{i}\left( t,\mathbf{x}^{\prime \prime }\right) =\delta
_{y^{\prime \prime }}^{i}\mathcal{A}\exp \left[ \mathrm{i}\left( Kz^{\prime
\prime }-\omega _{\mathrm{e}}t\right) \right] ,  \label{sol0}
\end{equation}%
where we let $\mathbf{x}^{\prime \prime }$ refer to the coordinates in the 
\textit{EMW} frame. Then using this for Eq. (\ref{p1}), by straightforward
computation, we obtain the simplest perturbation solution: 
\begin{eqnarray}
&&\delta A_{[{h}]}^{i}\left( t,\mathbf{x}^{\prime \prime }\right) =2\left(
\omega _{\mathrm{e}}/\omega _{\mathrm{g}}\right) A_{\mathrm{o}}^{i}\left( t,%
\mathbf{x}^{\prime \prime }\right)  \notag \\
&&\times \left[ h_{+}\cos ^{2}\left( \theta ^{\prime \prime }/2\right) \cos
\left( 2\psi ^{\prime \prime }\right) \cos \left( \mathbf{k}^{\prime \prime }%
\mathbf{\cdot x}^{\prime \prime }-\omega _{\mathrm{g}}t\right) \right. 
\notag \\
&&\left. -h_{\times }\cos ^{2}\left( \theta ^{\prime \prime }/2\right) \sin
\left( 2\psi ^{\prime \prime }\right) \sin \left( \mathbf{k}^{\prime \prime }%
\mathbf{\cdot x}^{\prime \prime }-\omega _{\mathrm{g}}t\right) \right] ,
\label{sol1}
\end{eqnarray}%
where%
\begin{equation}
\mathbf{k}^{\prime \prime }\equiv \left( k\sin \theta ^{\prime \prime }\cos
\phi ^{\prime \prime },k\sin \theta ^{\prime \prime }\sin \phi ^{\prime
\prime },k\cos \theta ^{\prime \prime }\right) ,  \label{k2}
\end{equation}%
and the angles $\left\{ \phi ^{\prime \prime },\theta ^{\prime \prime },\psi
^{\prime \prime }\right\} $ refer to the Euler rotations between $\mathbf{x}%
^{\prime }$ in the GW frame and $\mathbf{x}^{\prime \prime }$ in the EMW
frame, 
\begin{equation}
\mathbf{x}^{\prime }=\mathbf{R}\left( \phi ^{\prime \prime },\theta ^{\prime
\prime },\psi ^{\prime \prime }\right) \mathbf{x}^{\prime \prime }\mathbf{.}
\label{eu2}
\end{equation}

The above results can be extended to obtain the solutions for a general
case, in which light propagates along an arbitrary direction; rather than
along a single axis, e.g., the $z^{\prime \prime }$-axis. For this purpose,
one can consider $\mathbf{x}^{\prime \prime }=\left( x^{\prime \prime
},y^{\prime \prime },z^{\prime \prime }\right) $ in the EMW frame as rotated
from $\mathbf{x}=\left( x,y,z\right) $ in the detector frame, in which light
is seen to propagate along an arbitrary direction, as resulted from the
rotations of the $z^{\prime \prime }$-axis. The relation between the two
frames can be expressed by Euler angle rotations \cite{Goldstein,Rakhmanov};
but only with the direction angles $\left\{ \phi _{\star } ,\theta _{\star } \right\} $ in
spherical coordinates, without the polarization-ellipse angle: 
\begin{equation}
\mathbf{x}^{\prime \prime }=\mathbf{R}\left( \phi _{\star } ,\theta _{\star } \right) \mathbf{x,}
\label{eui}
\end{equation}%
where 
\begin{equation}
\mathbf{R}\left( \phi _{\star } ,\theta _{\star } \right) =\mathbf{R}_{\mathrm{II}}\left( \theta _{\star }
\right) \mathbf{R}_{\mathrm{I}}\left( \phi _{\star } \right) ,  \label{Ri}
\end{equation}%
with 
\begin{equation}
\mathbf{R}_{\mathrm{I}}\equiv \left[ 
\begin{array}{ccc}
\cos \phi _{\star } & \sin \phi _{\star } & 0 \\ 
-\sin \phi _{\star } & \cos \phi _{\star } & 0 \\ 
0 & 0 & 1%
\end{array}%
\right] ,\mathbf{R}_{\mathrm{II}}\equiv \left[ 
\begin{array}{ccc}
\cos \theta _{\star } & 0 & -\sin \theta _{\star } \\ 
0 & 1 & 0 \\ 
\sin \theta _{\star } & 0 & \cos \theta _{\star }%
\end{array}%
\right] .  \label{Ri12}
\end{equation}

Now, by means of Eq. (\ref{eui}) and the invariance relation 
\begin{equation}
\mathbf{K}^{\prime \prime }\mathbf{\cdot x}^{\prime \prime }=\mathbf{K\cdot x%
}\Leftrightarrow Kz^{\prime \prime }=K_{x}x+K_{y}y+K_{z}z,  \label{tr}
\end{equation}%
where $\mathbf{K}^{\prime \prime }=\left( 0,0,K\right) $ and $\mathbf{K}%
=\left( K_{x},K_{y},K_{z}\right) $ with $K=\sqrt{%
K_{x}^{2}+K_{y}^{2}+K_{z}^{2}}=\omega _{\mathrm{e}}/c$, one can express $%
\phi _{\star } $, $\theta _{\star } $ in terms of $K_{x}$, $K_{y}$, $K_{z}$: 
\begin{equation}
\sin \theta _{\star } \cos \phi _{\star } =\frac{K_{x}}{K},\sin \theta _{\star } \sin \phi _{\star } =\frac{K_{y}}{K}%
,\cos \theta _{\star } =\frac{K_{z}}{K}.  \label{euK}
\end{equation}%
Based on these, one can convert $\mathbf{R}\left( \phi _{\star } ,\theta _{\star } \right)
\rightarrow \mathbf{T}\left( K_{x},K_{y},K_{z}\right) $ and rewrite Eq. (\ref%
{eui}) as%
\begin{equation}
\mathbf{x}^{\prime \prime }=\mathbf{T}\left( K_{x},K_{y},K_{z}\right) 
\mathbf{x,}  \label{tr1}
\end{equation}%
where 
\begin{equation}
\mathbf{T}=\left[ 
\begin{array}{ccc}
\frac{K_{x}K_{z}}{K\sqrt{K_{x}^{2}+K_{y}^{2}}} & \frac{K_{y}K_{z}}{K\sqrt{%
K_{x}^{2}+K_{y}^{2}}} & -\frac{\sqrt{K_{x}^{2}+K_{y}^{2}}}{K} \\ 
-\frac{K_{y}}{\sqrt{K_{x}^{2}+K_{y}^{2}}} & \frac{K_{x}}{\sqrt{%
K_{x}^{2}+K_{y}^{2}}} & 0 \\ 
\frac{K_{x}}{K} & \frac{K_{y}}{K} & \frac{K_{z}}{K}%
\end{array}%
\right] .  \label{T}
\end{equation}%
The inverse transformation of (\ref{tr1}) is expressed by 
\begin{equation}
\mathbf{x}=\mathbf{T}^{-1}\left( K_{x},K_{y},K_{z}\right) \mathbf{\mathbf{x}%
^{\prime \prime },}  \label{tr2}
\end{equation}%
where $\mathbf{T}^{-1}$ is given by $\mathbf{T}^{\mathrm{T}}$, the transpose
of $\mathbf{T}$.

As seen in Section \ref{maxwell}, one can consider $\mathbf{x}^{\prime }$ in
the GW frame as rotated from $\mathbf{x}$ in the detector frame through the
Euler angles $\left\{ \phi ,\theta ,\psi \right\} $. Thus, combining Eq. (%
\ref{eu}) with Eq. (\ref{tr2}), and then comparing this with Eq. (\ref{eu2}%
), we find the relation:%
\begin{equation}
\mathbf{R}\left( \phi ^{\prime \prime },\theta ^{\prime \prime },\psi
^{\prime \prime }\right) =\mathbf{R}\left( \phi ,\theta ,\psi \right) 
\mathbf{T}^{-1}\left( K_{x},K_{y},K_{z}\right) .  \label{RRT}
\end{equation}%
That is, using Eqs. (\ref{R}), (\ref{R123}) and (\ref{T}) for this, one can
express $\phi ^{\prime \prime }$, $\theta ^{\prime \prime }$, $\psi ^{\prime
\prime }$ in terms of $\phi $, $\theta $, $\psi $ and $K_{x}$, $K_{y}$, $%
K_{z}$.

By means of Eqs. (\ref{tr1}), (\ref{tr2}) and (\ref{RRT}), one can transform
the solutions $A_{\mathrm{o}}^{i}\left( t,\mathbf{x}^{\prime \prime }\right) 
$ and $\delta A_{[{h}]}^{i}\left( t,\mathbf{x}^{\prime \prime }\right) $ in
the EMW frame, in which light propagates along the $z^{\prime \prime }$%
-axis, as given by Eqs. (\ref{sol0}) and (\ref{sol1}) respectively, to the
solutions for a general case, $A_{\mathrm{o}}^{i}\left( t,\mathbf{x}\right) $
and $\delta A_{[{h}]}^{i}\left( t,\mathbf{x}\right) $ in the detector frame,
in which light propagates along an arbitrary direction of $\mathbf{K}=\left(
K_{x},K_{y},K_{z}\right) $, as given by Eqs. (\ref{sol3}) and (\ref{sol4})
respectively.




\end{document}